\documentclass[pdflatex,sn-mathphys-num]{sn-jnl}
\usepackage[T1]{fontenc}
\usepackage[utf8]{inputenc}

\usepackage{graphicx}%
\usepackage{multirow}%
\usepackage{amsmath,amssymb,amsfonts}%
\usepackage{amsthm}%
\usepackage{mathrsfs}%
\usepackage[title]{appendix}%
\usepackage{xcolor}%
\usepackage{textcomp}%
\usepackage{manyfoot}%
\usepackage{booktabs}%
\usepackage{algorithm}%
\usepackage{algorithmicx}%
\usepackage{algpseudocode}%
\usepackage{listings}%
\usepackage{longtable}
\usepackage{rotating}

\usepackage{dcolumn}

\DeclareMathOperator{\sgn}{sgn}

\begin{document}

\title[Phase-Space Structure]{Discrete System Phase Space Modification by Vanishing Noise}%

\author*[1,2]{\fnm{Franco} \sur{Bagnoli}} \email{franco.bagnoli@unifi.it}
\author*[1,2]{\fnm{Michele} \sur{Baia}} \email{michele.baia@unifi.it}
\author*[1]{\fnm{Tommaso} \sur{Matteuzzi}} \email{tommaso.matteuzzi@unifi.it}
    
\affil[1]{
 \orgdiv{Dept.of Physics and Astronomy and CSDC}, \orgname{University of Florence}, \orgaddress{\street{via G. Sansone 1}, \postcode{50019} \city{Sesto Fiorentino},  \country{Italy}}}
\affil[2]{\orgname{INFN}, \orgdiv{Sect. Florence}, \orgaddress{\street{via G. Sansone 1}, \postcode{50019} \city{Sesto Fiorentino},  \country{Italy}}}

\abstract{
We investigate two classes of discrete systems: the Hopfield model with deterministic update dynamics and elementary cellular automata. We obtain a sampling of the attractor space of the Hopfield model and the complete phase space of all minimal elementary cellular automata for small lattice sizes. Starting from the maximal entropy distribution (all configurations equiprobable), we show how the dynamics affects this distribution. 
We then investigate how a vanishing noise alters this phase space, connecting attractors and modifying the asymptotic probability distribution. For both the Hopfield model and Cellular Automata, this modification does not always decrease the entropy.}

\maketitle

\catcode`\"=\active
\def"#1"{``#1''}

\let\at=@
\catcode`\@=\active
\def@#1{\ifmmode\boldsymbol{#1}\else\at#1\fi}
\newcommand{\De}[3][]{\frac{\partial^{#1} #2}{\partial {#3}^{#1}}}

\section{Introduction\label{sec:intro}}
We are interested in studying the phase space structure of discrete dynamical systems and its behavior under the action of a vanishing noise. We focus on two types of discrete systems: elementary cellular automata (CA) ~\cite{wolfram1983statistical} and a serial (deterministic) version the Hopfield network model~\cite{Hopfield1982,Abe1989,ramsauer2020hopfield}. The present paper is an extended version of a previous investigation limited to CA~\cite{Bagnoli2024}.

The Hopfield network was introduced as a model for associative memories. Giving a number of patterns (or configurations), there is a procedure for trying to make them attractors (fixed points) of the evolutionary dynamics of the model. Once stored, any initial configuration within the basin of attraction of a given pattern will eventually evolve toward it. Since the attractors of the Hopfield model  are all fixed points, the phase space of this model has a simple structure, which can be visualized as an energy landscape where trajectories descend toward local minima (not moving on a flas energy surface).

On the other hand, cellular automata are systems showing a much richer phase structure, see, for example, the series of proceedings of the ACRI conference to get a wide scenario of this subject~\cite{ACRI}. 

Deterministic CA have been introduced as discrete dynamical systems by S. Wolfram~\cite{wolfram1983statistical} and then deeply studied by A. Wuensche~\cite{wuensche1992global,wuensche1994complexity,Wuensche1999Classifying}.
A given CA dynamics connects configurations, originating trajectories. Since the evolution is deterministic, we can have joining trajectories, but not separation among them. For finite lattices, trajectories, after a transient, always end into a cycle or a fixed point (which is a cycle with period equal to one), that constitutes the only attractors. Similarly to what happens in the Hopfield model, the phase-space of a given CA rule is partitioned into basins of attraction. This characterization is not new at all, it has been studied for instance in Ref.~\cite{wuensche1992global,wuensche1994complexity}.

CA rules can be classified according on the pattern they generate,  or according to the distribution and character (cycle length) of the attractors, the size of their basins and their transients~\cite{wolfram1983statistical,PackardCAstructure}; many other classification techniques have been defined~\cite{Wuensche1999Classifying,MartinezECAClassification}. 

Another way of looking at the dynamics properties of a discrete dynamical system is that of measuring the stability of a trajectory with respect to some perturbation or noise, by observing the spreading of a damage~\cite{Stauffer1988} and defining indicators like Lyapunov exponents~\cite{Bagnoli1992,Shereshevsky1992,Baetens-LyapMultiStateCA}  or similar ones~\cite{LypunovProfile}.

One of the investigations reported in this paper is that of evaluating the probability distribution of configurations starting from a maximum-entropy distribution (i.e., random initial conditions) for the Hopfield model and all the minimal elementary CA rules and lattice sizes up to 17 sites. 

Our main goal is to examine how the phase space structure evolves under the action of a vanishing noise, i.e., a noise that is only occasionally present. This is particularly intriguing since, in nature, even  systems like genetic replication and proteoin synthesis, which are inherently robust to small noise due to their discrete dynamics, can still experience occasional perturbations that influence their behavior. 
In particular, the effect of noise in neural networks has been for instance examined in Ref.~\cite{Su2019}, where it has been shown that even a single-pixel modification can drastically alter the class an image belongs to. 

We shall show how a vanishing noise connects attractors, therefore determining a Markov process among them; the probability distribution of configurations forming the attractors is thus modified by this noise, and the corresponding entropy changes, in some cases even diminishing with the noise. 

The outline of this paper is the following. 
In Section~\ref{sec:definitions} we give some definition, then we present the phase space of the Hopfield  and of elementary CA model, their attractors and relative basins in Section~\ref{sec:Dynamics}. In Section~\ref{sec:attractorStability} we show how attractors are connected by a vanishing noise and how the phase-space probability distribution is modified by it. Conclusions are drawn in the last section.

\section{Definitions}\label{sec:definitions}

\subsection{Cellular automata}
Deterministic cellular automata are completely discrete dynamical  systems, defined on a lattice of cells, which can take a finite number of states. The evolution of the state of a cell is given by a function of its neighborhood, i.e., of the state of the cells connected to it with an incoming link. 

In order to be more specific, let us define a network though an adjacency matrix $a$, where  $a_{ij}=1$ if cell $j$ is connected (i.e., sends information) to cell $i$ and zero otherwise. In general, regular lattices with periodic boundary conditions are used, for which the adjacency matrix is shift-invariant (circulant). In particular, for elementary CA, a cell is connected to the cell itself and to its two nearest neighbors, i.e., the matrix $a$ is tri-diagonal. 

Let us denote by $s_i(t)$ the state of cell $i$ at time $t$. In order to be more concise, the index $t$ will be considered implicit, and we shall write $s_i'\equiv s_i(t+1)$. Similarly, the state of cells in the neighborhood of cell $i$ at time $t$ will be denoted as $v_i\equiv v_i(t)=\{s_j:a_{ij}=1\}$. ECA are Boolean CA, i.e., $s_i\in\{0,1\}$ and the neighborhood of a cell $i$ is $v_i=\{s_{i-1},s_i,s_{i+1}\}$, can also be read as a number in base-two representation, $v_i=4s_{i+1}+2 s_i+s_{i-1}$; $0\le v_i\le 7$.

\begin{table}[t]
    \caption{Look-up table of some CA rules. Here $s_-,s,s_+$ stands for $s_{i-1}, s_i, s_{i+1}$ and \textbf{R n} stands for rule $n$.}
    \label{tab:look}
    \begin{tabular}{c|c|c|c|c|c|c|c|c}
         $\mathbf{s_-,s,s_+}$& \textbf{R 1} & \textbf{R 164} & \textbf{R 232} & \textbf{R 11} & \textbf{R 30} & \textbf{R 150} & \textbf{R 204}& \textbf{R 110}\\
        \hline
\textbf{0,0,0}&1&0&0&1&0&0&0&0\\
\textbf{0,0,1}&0&0&0&1&1&1&0&1\\
\textbf{0,1,0}&0&1&0&0&1&1&1&1\\
\textbf{0,1,1}&0&0&1&1&1&0&1&1\\
\textbf{1,0,0}&0&0&0&0&1&1&0&0\\
\textbf{1,0,1}&0&1&1&0&0&0&0&1\\
\textbf{1,1,0}&0&0&1&0&0&0&1&1\\
\textbf{1,1,1}&0&1&1&0&0&1&1&0\\
    \end{tabular}
\end{table}

\begin{figure}[t]
    \centering
    \begin{tabular}{cccc}
        Rule 1 & Rule 11 & Rule 30 & Rule 110\\
        \includegraphics[width=0.21\textwidth]{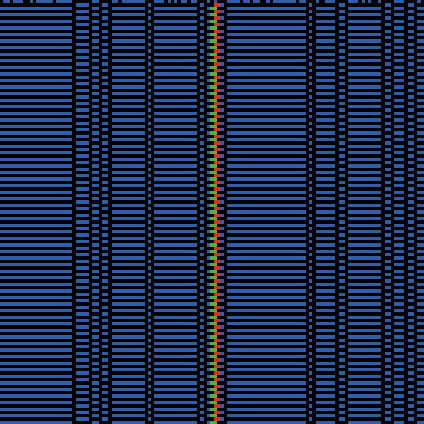} & \includegraphics[width=0.21\textwidth]{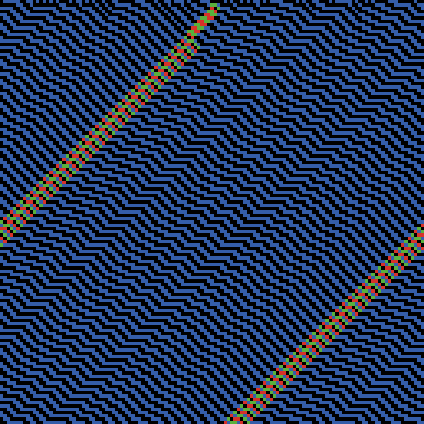} & \includegraphics[width=0.21\textwidth]{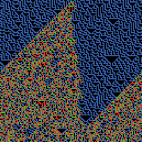} & 
        \includegraphics[width=0.21\textwidth]{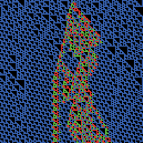}\\
        Rule 150 & Rule 164& Rule 204 & Rule 232\\
          \includegraphics[width=0.21\textwidth]{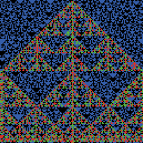} & \includegraphics[width=0.21\textwidth]{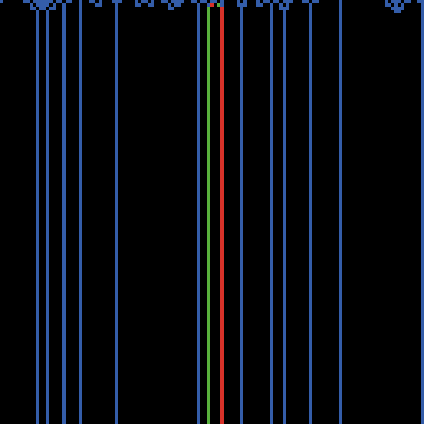} & \includegraphics[width=0.21\textwidth]{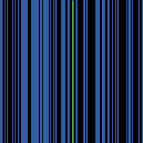} & \includegraphics[width=0.21\textwidth]{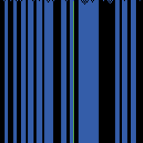} \\
    \end{tabular}
    \caption{Time evolution of some rules starting from a random initial configuration ($s_i(t)$) and the evolution of an initial single defect ($S_i(t)$). Time runs from top to down. Color code: black: $s_i(t)=S_i(t)=0$; blue: $s_i(t)=1$, $S_i(t)=0$; green: $s_i(t)=S_i(t)=1$; red: $s_i(t)=0$, $S_i(t)=1$.}
    \label{fig:evolv}
\end{figure}

The evolution of the state of a cell is given by a function of the neighborhood $f(v)$, which is applied in parallel to all cells, $s_i'=f(v_i)$.

Since the neighborhood $v$ can take only a finite number of values, the function $f$ can be seen as a look-up table. Therefore, there are $2^8=256$ possible Boolean functions of three inputs, and each function is specified by listing the 8 values corresponding to $v=0,\dots,7$. 

Reading again this list of values as a number in base-2 representation, $(f(1,1,1),f(1,1,0),\dots$ $,f(0,0,0))$ we get the Wolfram notation for ECA~\cite{wolfram1983statistical}, as illustrated for some rules in Table~\ref{tab:look}. By exploiting left-right and 0-1 symmetries, one can reduce the number of independent rules to 88, called the "minimal" rules, listed in Table~\ref{tab:attractors}.

The time evolution of some rules is reported in Fig.~\ref{fig:evolv}. Rule 1, 164, 232, 204 are typical class-2 rules, quickly falling into a fixed point or a cycle of small period. This implies that their phase-space is partitioned into many attractors of short period. Also, rule 11 is considered class-2 since it does not exhibit interesting patterns, although in this case the cycles are not so many and periods depend on lattice size. 
Rule 204 is the identity. Rule 232 is the majority rule. 
Rules 22, 30 and 150 are typical class-3, i.e., chaotic, and indeed for these rules an initial damage tends to spread (sensitivity to initial configuration). 
Rule 110 is class-4, capable of universal computing~\cite{cook}, exhibiting gliders and traveling structures.

\subsection{Deterministic Hopfield model}

The Hopfield model can be defined in a similar way. Let $N$ the number of cells in the system, the state $\sigma_i(t)$ of a cell is now in the set $\{-1; 1\}$. The state of a cell evolves following a rule that depends on the value of the local field, $h_i$,
\begin{equation}\label{eq:h}
    h_i =\sum_{i=1}^N w_{ij}\sigma_j(t),
\end{equation}
as
\begin{equation}\label{eq:sigma}
    \sigma'_i = \begin{cases}
    \sigma_i(t) & \text{if $h_i=0$}\\
    \sgn (h_i(t)) &\text{otherwise},
    \end{cases}
\end{equation}
where the interaction matrix $w$ is defined by the $K$ stored patterns $\xi_i^{(k)}\in \{-1; 1\}$, $i=1,\dots,N$, $k=1,\dots,K$,
\begin{equation}\label{eq:w}
    w_{ij} = \sum_{k=1}^K \xi_i^{(k)}\xi_j^{(k)}.
\end{equation}

The application of Eq.~\eqref{eq:sigma} is performed sequentially, since with the parallel application of this evolution rule does not guarantee that the patterns $\xi_i^{k}$ are minima of energy, even for $K=2$, and the system may oscillate. 

Instead of updating the cells in a random order, as is typically done, we update them sequentially in a fixed order, $i=1,\dots,N$, ensuring a fully deterministic algorithm. The field $h_i$ is recomputed for each value of $i$ using the updated values of $\sigma_j$, $j<i$.

The Hopfield system is completely characterized by the energy function:
\[
H = -\sum_i\sum_j w_{ij}\sigma_i\sigma_j
\]
and the state of the system evolves toward minimums of $H$ (which are fixed points). If the number $K$ of stored patterns is much less than $N$, $\xi_i^{(k)}$ and their opposites (replacing $-1$ with $1$ and vice versa) are the only attractors of the dynamics. However, as the number of stored patterns increases, additional spurious attractors emerge, unrelated to the $\xi_i^{(k)}$, causing the network to lose the ability to retrieve the original patterns.

\begin{figure}[t]
    \centering
    \includegraphics[width=0.75\linewidth]{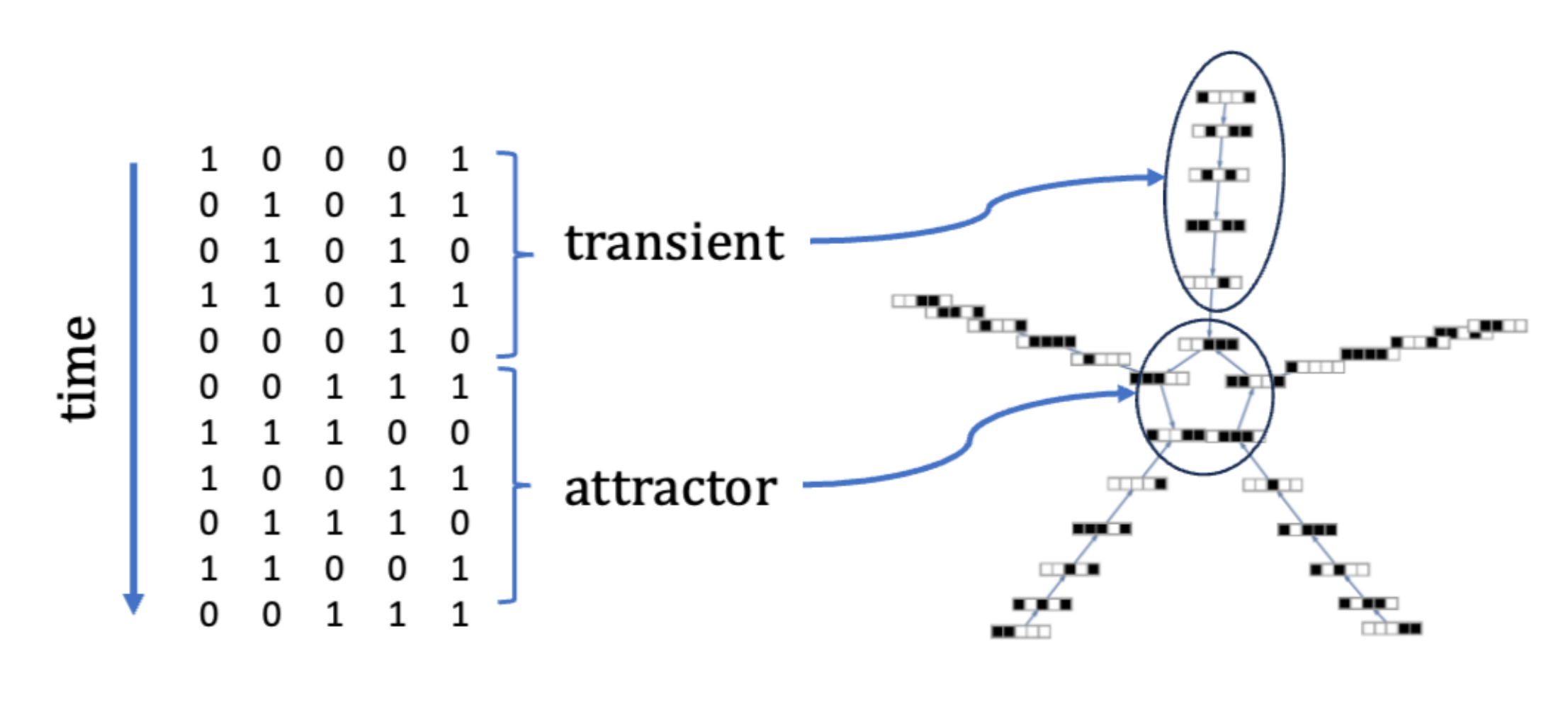}
    \caption{Transients and attractors for a basin of Rule 30, $L=5$. For this length, Rule 30 has another attractor, the fixed point $00000$ whose basin also includes the configuration $11111$. }
    \label{fig:transient-attractor}
\end{figure}

\begin{figure}[t]
    \centering
    \begin{tabular}{|c|c|c|c|}
    \hline        
     Rule 1 & Rule 11 & Rule 30 & Rule 110\\
    \hline 
        \includegraphics[width=0.22\textwidth]{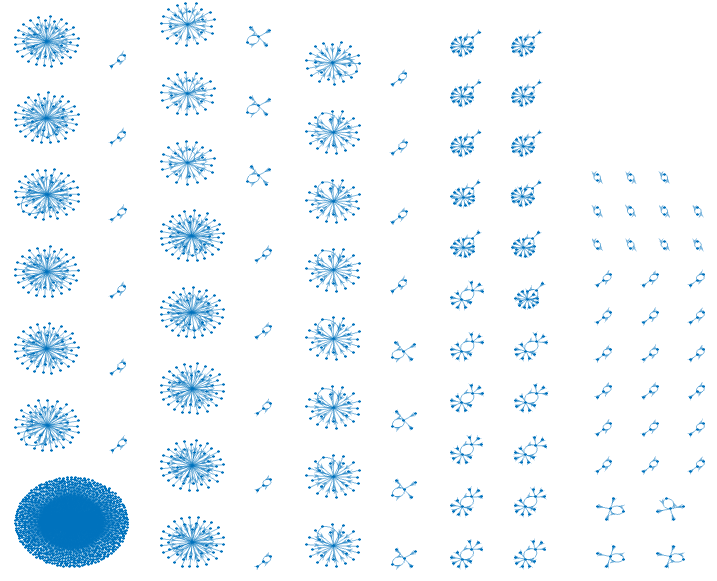} & \includegraphics[width=0.22\textwidth]{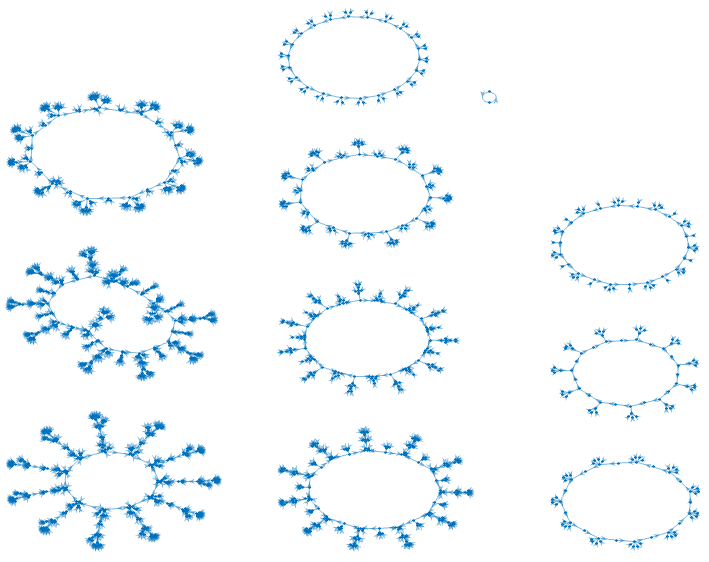} & \includegraphics[width=0.22\textwidth]{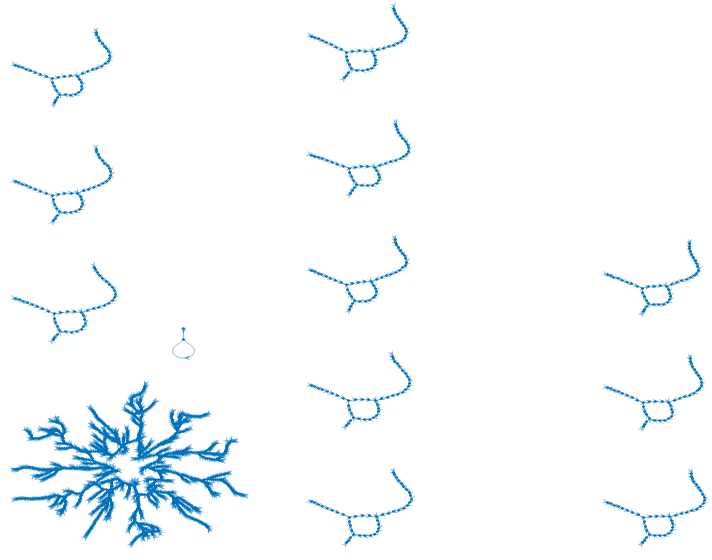} &\includegraphics[width=0.22\textwidth]{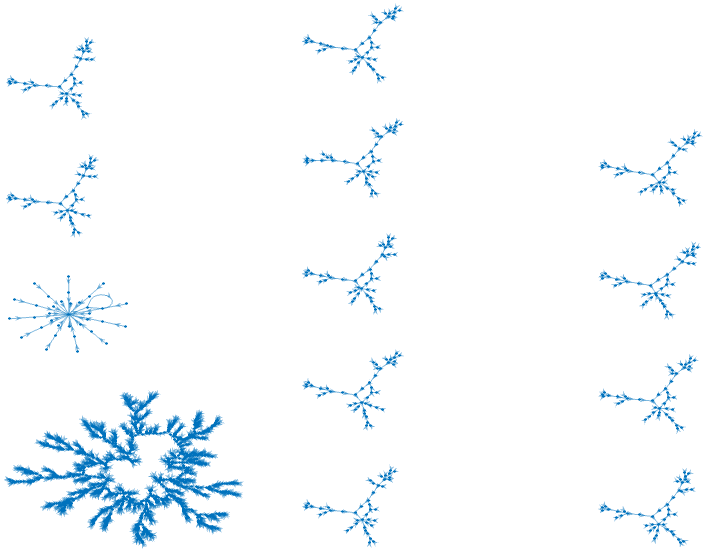}\\
        \hline
        \hline 
        Rule 150 & Rule 164& Rule 204 & Rule 232\\
        \hline 
          \includegraphics[width=0.22\textwidth]{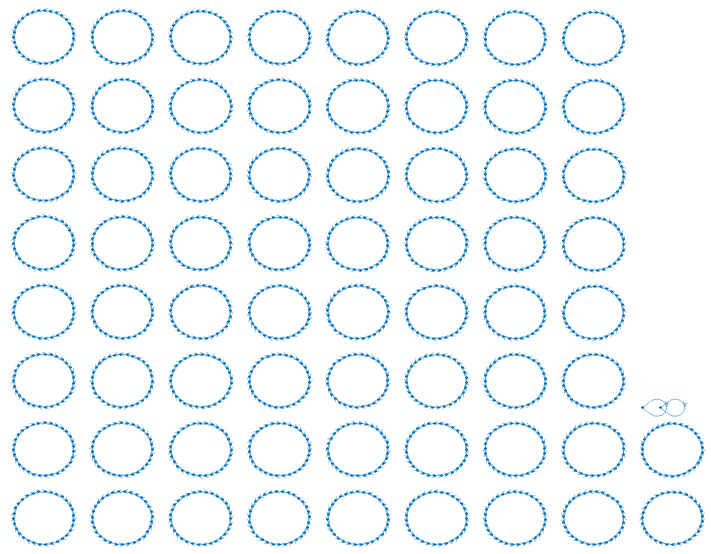} & \includegraphics[width=0.22\textwidth]{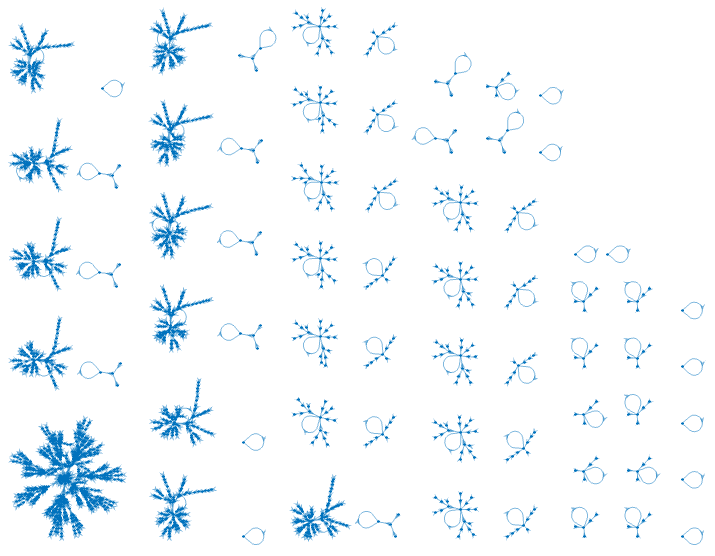} &
          \includegraphics[width=0.22\textwidth]{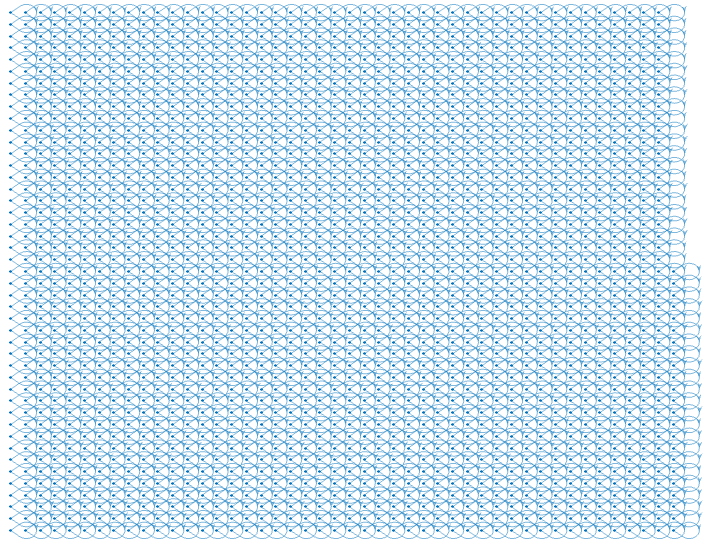} & \includegraphics[width=0.22\textwidth]{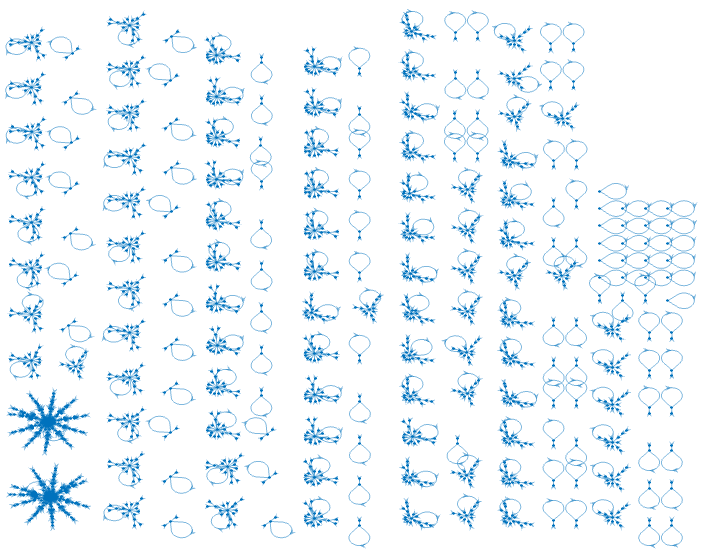} \\
          \hline 
    \end{tabular}
    \caption{Attractors of rules shown in Fig.~\ref{fig:evolv}. Depending on the rule the dynamics shows different behaviors; for example,  fixed point attractors (e.g., bottom left of rule 1), cycles with/without a basin of attraction (e.g., rule 30 and 150). }
    \label{fig:attractors}
\end{figure}

\section{Dynamics, phase space and attractors\label{sec:Dynamics}}

The dynamics of these systems can be described from two complementary perspectives. From a dynamical systems viewpoint, the focus is on evolution in terms of trajectories. Alternatively, to emphasize the evolution of the probability distribution over configurations, the system can be described as (degenerate) Markov process.

\subsection{Dynamical system description}

For both, CA and Hopfield model, one can consider the state of  the whole system as, 
\[
@s(t)\equiv \{s_i(t), i=1,\dots,L\}.
\]

Defining a global function, $@F$, so that 
\[
@s(t+1)=@F(@s(t)).
\]
(for CA, with appropriate boundary conditions) we have a discrete dynamical system. For a Boolean lattice of $L$ sites, $@s$ can take $2^L$ possible values. Let us denote a trajectory $\gamma$ the ordered set of configurations given by the dynamics, 
\[
\gamma\equiv \gamma(@s(0))=\{@s(0),@s(1),\dots\}.
\]

Since the evolution is deterministic and the number of states (on a finite lattice) is finite, a trajectory always ends in a cycle or in a fixed point (which is cycle of length $1$) as for the Hopfield model.

A cycle $\alpha$ is a trajectory $\alpha=\{@s(1),@s(2),\dots,@s(k)\}$ such that $@s(1)=@F(@s(k))$. The length of the cycle is indicated by $c(\alpha)$.  

A generic trajectory is represented by a (possible empty) transient $\tau$ followed by one of the configurations in an attractor $\alpha$: $\gamma(@s(0)) =\{\tau(@s(0)),\alpha,\alpha,\dots\}$ where $\tau$ is the minimal one, i.e., it has no overlap with $\alpha$, see Fig.~\ref{fig:transient-attractor} for an illustration.

The basin $\beta(\alpha)$ of an attractor $\alpha$ is given by all configurations $@s$ such that $\gamma(@s)=\{\tau(@s),\alpha\}$. The size of the basin is indicated by $b(\alpha)$. We can also arbitrarily number the attractors with an index $k=1,\dots,n$, so we can speak of the length $c_k$ of the attractor $k$,  and of the size $b_k$ of its basin. Clearly, $\sum_k b_k=2^L$.

\begin{figure}[t]
    \centering
    \includegraphics[width=0.5\linewidth]{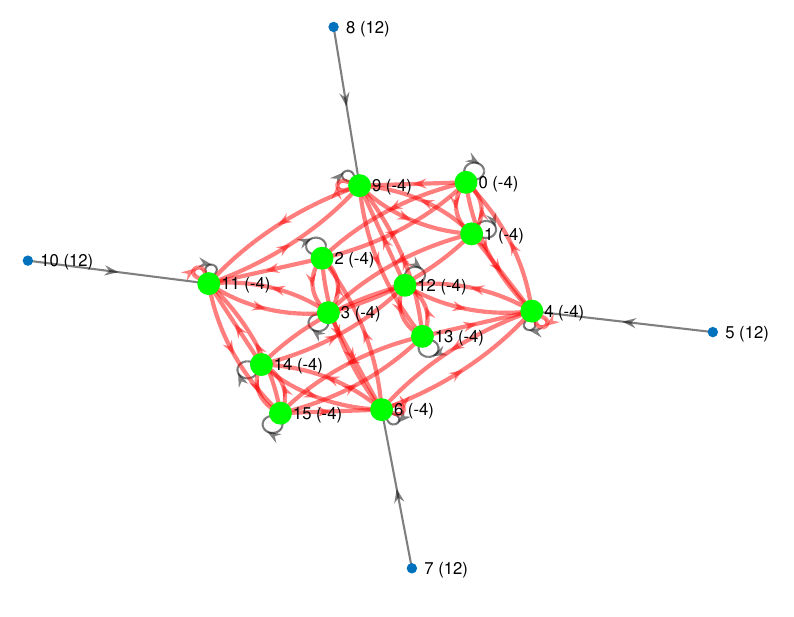}
    \caption{The ideal phase space of the Hopfield model with $L=4$ and $K=6$ stored patterns. In this case all stored patterns and their inverses are attractors of the dynamics (black arrows). The effect of noise (red arrows) is that of connecting all attractors.}
    \label{fig:HopfieldIdeal}
\end{figure}

\begin{figure}[t]
    \centering
\begin{tabular}{cc} 
(a) & (b)\\
\includegraphics[width=0.48\linewidth]{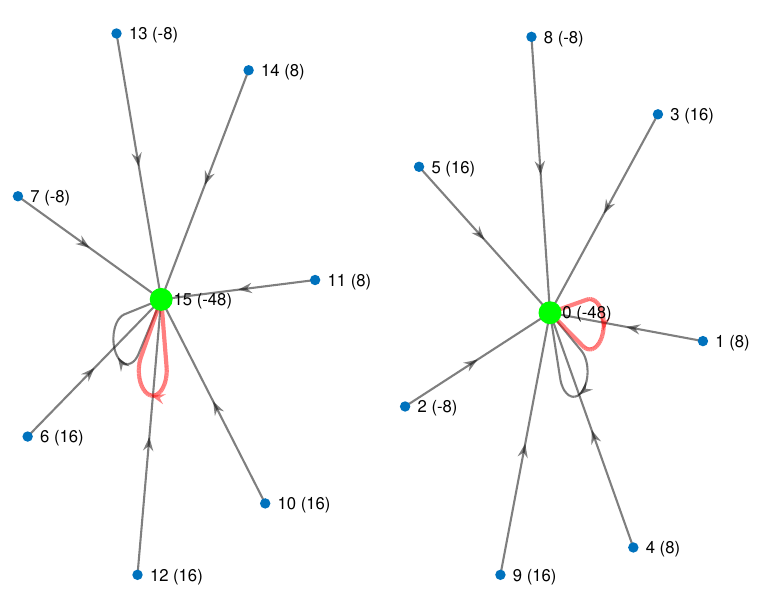} &
   \includegraphics[width=0.48\linewidth]{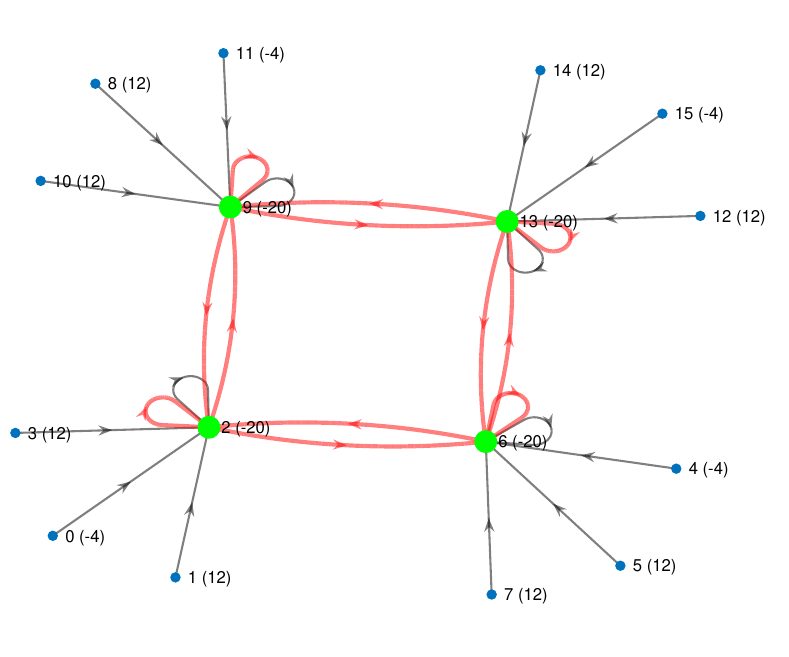}\\
(c) & (d)\\
\includegraphics[width=0.48\linewidth]{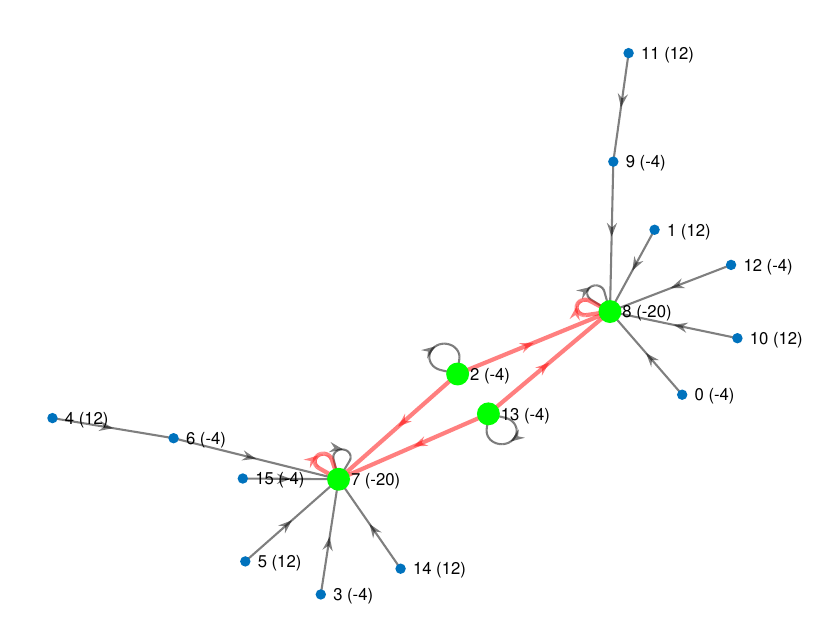} &\includegraphics[width=0.48\linewidth]{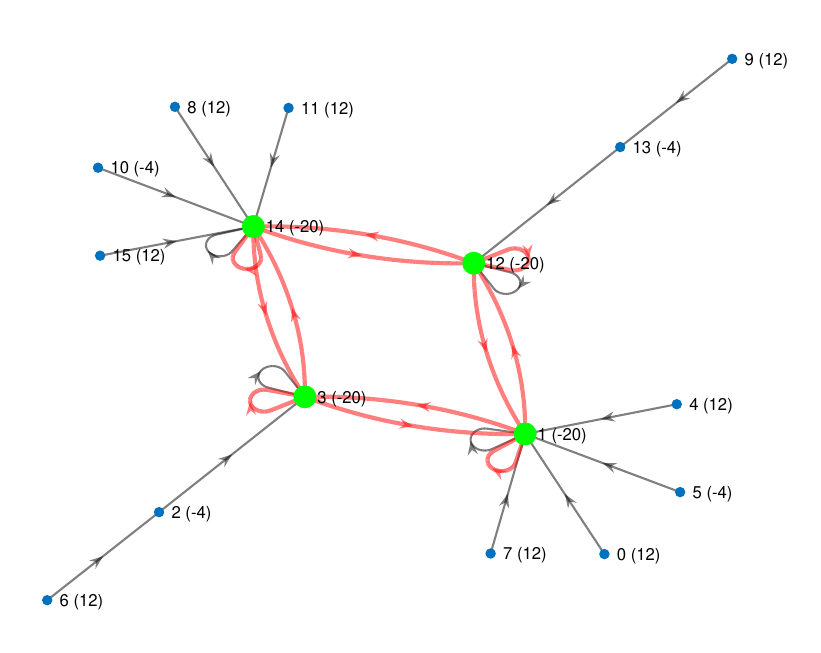}
 \end{tabular}
    \caption{Instances of the phase space of the Hopfield model with dimension $L=4$ and number of stored patterns $K=6$. In black the transitions given by the dynamics, Eq.~\eqref{eq:sigma}, in red the transitions induced by the vanishing noise. Nodes are indicated by configuration (expressing the number in base-two  representation) and energy in parenthesis. The attractors are marked by green dots.  (a) Two resulting attractors, both stable with respect to the noise. The entropy is $S=S^*=0.5$. (b)  Four  attractors, all with the same basin size. The noise connects attractors but does not affect the entropy ($S=S^*=0.25$). (c) Four attractors with different basin size. Two of them are unstable with respect to noise, which reduces the entropy from $S=0.386$ to $S^*=0.250$. (d) Four attractors, with different basin size. The entropy connects the attractors flattening the probability distributions. The entropy increases from $S=0.489$ to  $S^*=493$.}
    \label{fig:HopfieldNoise}
\end{figure}

\subsection{Markov process description}

Another approach to the same problem is the following. Let us define the connection matrix $M_{@s',@s}=M(@s'|@s)= 1$ if $@s'=@F(@s)$ and zero otherwise (it is a $2^L\times2^L$ matrix). $M$ is a (degenerate) Markov matrix, with $\sum_{@s}M_{@s',@s}=1$ (actually, only one entry per columns of $M$ is one, all the rest is zero). If we start with a probability distribution of configurations $@P(@s,0)$, we have
\[
@P(@s', t+1)=\sum_{@S}M(@s'|@s)@P(@s,t).
\]

Starting with a delta, $@P(@s, 0)=[@s=@s(0)]$, we get a sequence of deltas corresponding to the trajectory starting from $@s(0)$. After a transient time, the evolution enters an attractor. The distributions corresponding to the configurations belonging to an attractor are eigenvectors of $M$. 

We are start from  the maximally entropic initial state $@P(@s,0)=1/2^{L}$, for which the entropy is $S_{\mathrm{MAX}}=L\ln(2)$. Iterating the evolution for a long time, we  get an asymptotic distribution $\overline{@P}$ whose entries are zero for the configurations belonging to the transient part of the attractors and different from zero for the configurations belonging to the attractors. If all the  configurations  of all the attractor have the same probability, the asymptotic probability is equal to $b_k/(2^L c_k)$ for the configurations belonging to attractor $k$. This is always the case for the Hopfield models, since in this case all attractors are fixed points, but this is not always so for CA. For instance, for rule 1 (see Fig.~\ref{fig:attractors}), there are period-two attractors whose configurations are populated with a very different probability. 

In the case of even probability on all the attractors, for each of them ($k$),  there are $c_k$ configurations, and the entropy $S$ induced by the dynamics (normalized with respect to the maximum entropy $S_{\mathrm{MAX}}$) is
\[
S(\overline{@P})=\frac{-1}{S_{\mathrm{MAX}}}\sum_{@s} \overline{@P}(@s)\ln(\overline{@P}(@s))=\frac{-1}{2^LL\ln(2)}\sum_{k=1}^n b_k \ln\left(\frac{b_k}{c_k 2^L}\right).
\]

 For CA, it may happen that the asymptotic distribution is oscillating, for instance rule 1 maps all local configurations but $(0,0,0)$ to  $0$, and the local configuration $(0,0,0)$ to  $1$. So, starting from a homogeneous configuration,  we have an alternation of deltas (corresponding to configurations $@0=(0,0,\dots,0)$ and $@1=(1,1,\dots,1)$). Oscillations are not possible in the Hopfield model due to the sequential update.

The dynamics induces a production of information, that is, a reduction of entropy, which is related to the presence of transient states. One can have a visual representation of this contraction by looking at Fig.~\ref{fig:attractors}. We can see that in many cases there are transients, but this is not always true. For instance, rule 204 (the identity) only has fixed points, while rule 150 has many cycles and two fixed points ($@0$ and $@1$) with no transients. 

The entropy $S$ for for all ECA rules and $L=17$ is reported in Table~\ref{tab:attractors}. Several example where discussed more extensively in our previous work \cite{Bagnoli2024}.

For the Hopfield model the scenario is similar, except that the attractors are all fixed points. The ideal situation is that the $K$ stored patterns and their inverse are all (and the only) attractors of the dynamics, as shown in Fig.~\ref{fig:HopfieldIdeal} for $L=4$ and $K=6$, but this is a very uncommon scenario if the number of stored pattern exceed some percent of the number of neurons. The typical scenarios are that shown in Fig.~\ref{fig:HopfieldNoise}, with fewer attractors (not corresponding to stored patterjns) and larger basins.  

The values of the mean entropy for an ensemble of $1000$ Hopfield networks with $L=17$ and various number of stored patterns are shown In Fig.~\ref{fig:Hopfield}.

The computation of asymptotic probability distributions using the matrix $M$ (iterating or finding eigenvectors) is however computationally quite expensive. We gave a more effective procedure in the following section.

\subsection{Stability properties}

It is possible to define a maximum Lyapunov exponent related to a given trajectory of a discrete deterministic system, like for instance Cellular Automata~\cite{Bagnoli1992} or the deterministic Hopfield model. This Lyapunov exponent, similarly to what happens for continuous system, is positive for unstable  trajectories, in the sense that the smallest perturbation (one bit for discrete systems) can bring the system to another trajectory. The difference with respect to continuous system is that, even an unstable trajectory cannot be left, due to the discrete nature of our system. However, finite noise can make the system wander among trajectories, until possibly finding a stable one (negative Lyapunov exponent), robust with respect to such perturbations. 

For instance, the trajectory of the majority rule 232 reported in Fig.~\ref{fig:evolv} is unstable and shows a positive Lyapunov exponent. For this rule any block of at least two cells with the same value gives origin to a permanent "stripe". By adding a vanishing noise (i.e., at most one cell state flip per time step), the boundaries of a stripe can move. Upon meeting, two boundaries fuse, making the strip disappear, while a new strip cannot appear (unless the noise is so high that two neighboring defects are created at the same time). Finally, all stripes coalesce (for periodic boundary conditions, if boundaries are fixed a frustration may be present). The stable configuration in the presence of a vanishing noise are those composed by all zeros and all ones, which have a negative (minus infinity) Lyapunov exponent. 

These considerations illustrate the importance of examining the effects of a vanishing noise. It is possible to extend these concepts to any other deterministic dynamical system, such as the serial Hopfield model.

\begin{table}[h] 
\caption{Number of attractors $n$, the normalized entropy $S$ induced by dynamics , the normalized entropy $S_f$ assuming that the cycles are evenly populated and the normalized entropy $S^*$ induced by a vanishing noise. Computations for all minimal rules \textbf{R}, lattice size $L=17$, $n_\mathrm{MAX}=131072$.}\label{tab:attractors}
\centering
\tiny
\begin{tabular}{c|c|c|c|c||c|c|c|c|c}
\textbf{R} & $n$ &   $S$ & $S_f$ & $S^*$ & \textbf{R} & $n$ &  $S$ & $S_f$ & $S^*$ \\
\hline
$0$ & $ 1$ & $ 0.0$ & $ 0.0$ & $ 0.0$ & $ 56$ & $ 218$ & $ 0.613$ & $ 0.613$ & $ 0.605$ \\
$1$ & $ 1786$ & $ 0.434$ & $ 0.486$ & $ 0.634$ & $ 57$ & $ 15$ & $ 0.343$ & $ 0.343$ & $ 0.299$ \\
$2$ & $ 40$ & $ 0.485$ & $ 0.485$ & $ 0.538$ & $ 58$ & $ 20$ & $ 0.407$ & $ 0.407$ & $ 0.24$ \\
$3$ & $ 422$ & $ 0.699$ & $ 0.736$ & $ 0.766$ & $ 60$ & $ 260$ & $ 0.941$ & $ 0.941$ & $ 0.941$ \\
$4$ & $ 3571$ & $ 0.508$ & $ 0.508$ & $ 0.612$ & $ 62$ & $ 1186$ & $ 0.608$ & $ 0.611$ & $ 0.607$ \\
$5$ & $ 7158$ & $ 0.699$ & $ 0.736$ & $ 0.766$ & $ 72$ & $ 664$ & $ 0.323$ & $ 0.323$ & $ 0.0$ \\
$6$ & $ 48$ & $ 0.573$ & $ 0.577$ & $ 0.607$ & $ 73$ & $ 1361$ & $ 0.699$ & $ 0.71$ & $ 0.681$ \\
$7$ & $ 42$ & $ 0.496$ & $ 0.502$ & $ 0.059$ & $ 74$ & $ 101$ & $ 0.641$ & $ 0.642$ & $ 0.656$ \\
$8$ & $ 1$ & $ 0.0$ & $ 0.0$ & $ 0.0$ & $ 76$ & $ 31553$ & $ 0.85$ & $ 0.85$ & $ 0.851$ \\
$9$ & $ 24$ & $ 0.449$ & $ 0.45$ & $ 0.451$ & $ 77$ & $ 3571$ & $ 0.6$ & $ 0.6$ & $ 0.299$ \\
$10$ & $ 211$ & $ 0.678$ & $ 0.678$ & $ 0.687$ & $ 78$ & $ 120$ & $ 0.377$ & $ 0.377$ & $ 0.24$ \\
$11$ & $ 108$ & $ 0.579$ & $ 0.594$ & $ 0.299$ & $ 90$ & $ 4404$ & $ 0.941$ & $ 0.941$ & $ 0.941$ \\
$12$ & $ 3571$ & $ 0.678$ & $ 0.678$ & $ 0.687$ & $ 94$ & $ 1191$ & $ 0.561$ & $ 0.564$ & $ 0.526$ \\
$13$ & $ 120$ & $ 0.377$ & $ 0.377$ & $ 0.24$ & $ 104$ & $ 239$ & $ 0.208$ & $ 0.208$ & $ 0.0$ \\
$14$ & $ 120$ & $ 0.507$ & $ 0.507$ & $ 0.349$ & $ 105$ & $ 4370$ & $ 1.0$ & $ 1.0$ & $ 1.0$ \\
$15$ & $ 3856$ & $ 1.0$ & $ 1.0$ & $ 1.0$ & $ 106$ & $ 214$ & $ 0.698$ & $ 0.699$ & $ 0.701$ \\
$18$ & $ 120$ & $ 0.563$ & $ 0.578$ & $ 0.586$ & $ 108$ & $ 10966$ & $ 0.775$ & $ 0.776$ & $ 0.764$ \\
$19$ & $ 1786$ & $ 0.62$ & $ 0.63$ & $ 0.059$ & $ 110$ & $ 20$ & $ 0.494$ & $ 0.495$ & $ 0.493$ \\
$22$ & $ 52$ & $ 0.438$ & $ 0.44$ & $ 0.46$ & $ 122$ & $ 120$ & $ 0.584$ & $ 0.588$ & $ 0.588$ \\
$23$ & $ 1786$ & $ 0.6$ & $ 0.6$ & $ 0.059$ & $ 126$ & $ 120$ & $ 0.563$ & $ 0.578$ & $ 0.588$ \\
$24$ & $ 40$ & $ 0.544$ & $ 0.544$ & $ 0.535$ & $ 128$ & $ 2$ & $ 0.0$ & $ 0.0$ & $ 0.0$ \\
$25$ & $ 27$ & $ 0.451$ & $ 0.453$ & $ 0.441$ & $ 130$ & $ 41$ & $ 0.525$ & $ 0.525$ & $ 0.538$ \\
$26$ & $ 148$ & $ 0.782$ & $ 0.792$ & $ 0.809$ & $ 132$ & $ 3572$ & $ 0.6$ & $ 0.6$ & $ 0.624$ \\
$27$ & $ 422$ & $ 0.791$ & $ 0.801$ & $ 0.807$ & $ 134$ & $ 49$ & $ 0.549$ & $ 0.551$ & $ 0.578$ \\
$28$ & $ 630$ & $ 0.492$ & $ 0.501$ & $ 0.299$ & $ 136$ & $ 2$ & $ 0.0$ & $ 0.0$ & $ 0.0$ \\
$29$ & $ 15777$ & $ 0.863$ & $ 0.863$ & $ 0.857$ & $ 138$ & $ 837$ & $ 0.791$ & $ 0.791$ & $ 0.764$ \\
$30$ & $ 7$ & $ 0.791$ & $ 0.799$ & $ 0.799$ & $ 140$ & $ 3572$ & $ 0.678$ & $ 0.678$ & $ 0.687$ \\
$32$ & $ 1$ & $ 0.0$ & $ 0.0$ & $ 0.0$ & $ 142$ & $ 317$ & $ 0.572$ & $ 0.572$ & $ 0.393$ \\
$33$ & $ 1786$ & $ 0.626$ & $ 0.638$ & $ 0.664$ & $ 146$ & $ 121$ & $ 0.585$ & $ 0.587$ & $ 0.586$ \\
$34$ & $ 211$ & $ 0.678$ & $ 0.678$ & $ 0.687$ & $ 150$ & $ 8740$ & $ 1.0$ & $ 1.0$ & $ 1.0$ \\
$35$ & $ 429$ & $ 0.68$ & $ 0.689$ & $ 0.683$ & $ 152$ & $ 41$ & $ 0.515$ & $ 0.515$ & $ 0.535$ \\
$36$ & $ 664$ & $ 0.323$ & $ 0.323$ & $ 0.445$ & $ 154$ & $ 1688$ & $ 1.0$ & $ 1.0$ & $ 1.0$ \\
$37$ & $ 349$ & $ 0.531$ & $ 0.531$ & $ 0.517$ & $ 156$ & $ 631$ & $ 0.502$ & $ 0.502$ & $ 0.299$ \\
$38$ & $ 234$ & $ 0.726$ & $ 0.731$ & $ 0.738$ & $ 160$ & $ 2$ & $ 0.0$ & $ 0.0$ & $ 0.0$ \\
$40$ & $ 8$ & $ 0.001$ & $ 0.001$ & $ 0.0$ & $ 162$ & $ 212$ & $ 0.667$ & $ 0.667$ & $ 0.694$ \\
$41$ & $ 16$ & $ 0.457$ & $ 0.458$ & $ 0.467$ & $ 164$ & $ 665$ & $ 0.391$ & $ 0.391$ & $ 0.487$ \\
$42$ & $ 1857$ & $ 0.839$ & $ 0.839$ & $ 0.862$ & $ 168$ & $ 212$ & $ 0.03$ & $ 0.03$ & $ 0.0$ \\
$43$ & $ 316$ & $ 0.572$ & $ 0.572$ & $ 0.393$ & $ 170$ & $ 7712$ & $ 1.0$ & $ 1.0$ & $ 1.0$ \\
$44$ & $ 664$ & $ 0.528$ & $ 0.528$ & $ 0.538$ & $ 172$ & $ 704$ & $ 0.5$ & $ 0.5$ & $ 0.538$ \\
$45$ & $ 22$ & $ 1.0$ & $ 1.0$ & $ 1.0$ & $ 178$ & $ 1787$ & $ 0.6$ & $ 0.6$ & $ 0.299$ \\
$46$ & $ 40$ & $ 0.544$ & $ 0.544$ & $ 0.533$ & $ 184$ & $ 422$ & $ 0.572$ & $ 0.572$ & $ 0.567$ \\
$50$ & $ 1786$ & $ 0.6$ & $ 0.6$ & $ 0.299$ & $ 200$ & $ 14197$ & $ 0.699$ & $ 0.699$ & $ 0.0$ \\
$51$ & $ 65536$ & $ 1.0$ & $ 1.0$ & $ 1.0$ & $ 204$ & $ 131072$ & $ 1.0$ & $ 1.0$ & $ 1.0 $\\
$54$ & $ 124$ & $ 0.5$ & $ 0.501$ & $ 0.52$ & $ 232$ & $ 3572$ & $ 0.6$ & $ 0.6$ & $ 0.059$ \\
\end{tabular}
\end{table} 

\section{Numerical determination of attractors for small configurations\label{sec:numerics}}

The problem of finding the attractor and its basin starting from a given configuration was investigated using a preimage algorithm in Ref.~\cite{wuensche1992global}, which is much faster than investigating all the $2^L$ configurations for a lattice size $L$. However, since we are interested in enumerating all the attractors, we cannot avoid it. 

We can find all attractors and their characteristics for a small-size cellular automata, by enumerating all of them, following their evolution until entering a limit cycle, and numbering these cycles. Let $L$ be the size of CA. There are $2^L$ possible configurations $@s$, which can be read as a number $s$ in  base-two representation from 0 to $2^L-1$. Let $A$ be an array of size $2^L$ set to zero. The idea is to label each configuration in a trajectory $\{s(0), s(1), \dots\}$ with the same index $-k$, until we find a configuration $S$ which is already marked, i.e., $A(S)\neq 0$. If $A(S)=-k$ then we proceed by inverting the sign of all configurations in the trajectory until we find a positive entry, i.e., we have explored all the cycle. We then increment $k$.

If however $|A(S)|\neq k$, it means that the trajectory belongs to the basin of an already encountered attractor, so we restart from $s(0)$ and we change all $k$ with $|A(S)|$. We do not increase $k$ after this phase. 

When there are no more configurations we have found all attractors and basins and we set their number to $n=k-1$. Then we can compute, for each attractor $k\in\{1,\dots,n\}$, the length of its cycle (the number of entries $A(s)= k$) and its size (the number of entries $|A(s)|=k$).

\begin{figure}[t]
    \centering
    \includegraphics[width=1\linewidth]{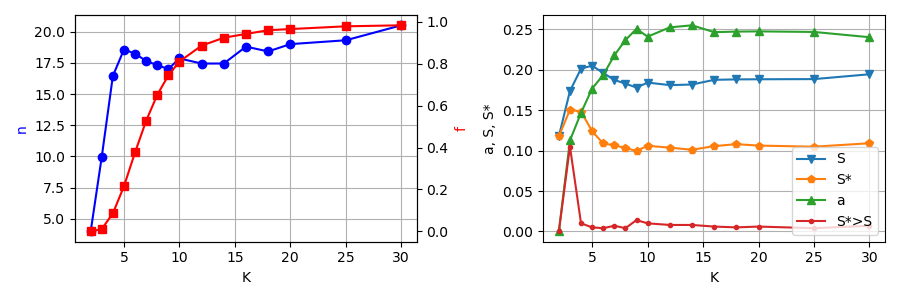}
    \caption{Hopfield model with $L=17$ averaged over $1000$ repetitions; $K$ is the number of stored patterns, $n$ is the number of actual attractors; $f$ is the fraction of attractors different from the stored patterns; $a$ measures the asymmetry of the noise-induces transitions; $S$ is the entropy induced by the dynamics; $S^*$ is the entropy with noise; $S^*>S$ is the fraction of times noise entropy is grater than dynamics entropy.}
    \label{fig:Hopfield}
\end{figure}

\section{Stability of attractors with respect to a vanishing noise\label{sec:attractorStability}}

The primary focus of our discussion is the effect of vanishing noise, that is, the occasional flipping of a single site’s value, as previously illustrated for rule 232.

Since the noise is vanishing, we assume that the system has already reached an attractor. A noise-induced flip causes a transition from the current configuration to a perturbed one. This new configuration may either remain within the basin of the same attractor or belong to that of a different attractor. Thus, the key effect of vanishing noise is the potential connection between attractors. This happens if the basin of one attractor is at distance one (minimal noise) from one configuration belonging to another attractor. 

Even if attractors are isolated, the effect of noise is that of flattening the probability distribution of configurations belonging to the attractor itself, so from what concerns the transition induced by the noise on we assume that the probability of all configurations belonging to attractor $k$ is $b_k/(c_k 2^L)$.  

It is possible to build a matrix $W_{ij}\equiv W(i|j)$ which counts, for all configurations belonging to cycle $j$, and for all possible one-site perturbations, how many of them will belong to the basin of attractor $i$. 

Normalizing it over columns (index $i$), one gets a Markov matrix  which gives the probability of going from attractor $j$ to attractor $i$ under the influence of a vanishing noise. 

So, beyond the entropy reduction given by dynamics, we can have an entropy variation due to this noise. We shall denote the normalized entropy of the probability distribution $\overline{@Q}$, obtained by iterating $@Q' =W@Q$, by the symbol $S^*$. 

What is interesting, is that this vanishing noise can act in  different ways on CA rules and on the Hopfield model, leading to an increase or a decrease of the entropy depending on the phase space structure of the system. 

Results are reported in Table~\ref{tab:attractors} for all the elementary CA rule and in Fig.~\ref{fig:Hopfield} for the Hopfield model for $L=17$ and different number $K$ of stored patterns.

\section{Noise effect}

The effect of noise depends on the stability of attractor trajectories. Here, we recall the results for  CA (already reported in \cite{Bagnoli2024}) and we discuss the effects of noise in the Hopfield model. 

The comparison of $S$ and $S^*$ for all elementary CA rules is reported in Table~\ref{tab:attractors}, along with the number of attractors. The effect of noise is clearly zero for "contracting" rules like rule 0, 8, etc. For these rules there is only one attractor, a stable fixed point (with Jacobian equal to zero), so nothing happens and $S^*=S=0$. The same happens for "flat" rules like the identity 204 or the shift 170. In these cases, the Jacobian is diagonal (eventually circularly shifted by one) and therefore the perturbation is simply maintained, so $S^*=S=1$, irrespective of noise. A similar result occurs for the linear rule 150 where the effect of noise propagates on all the lattice, but, since there is no preferred attractor, it has no influence on the probability distribution. 

In general, the statistical properties of "chaotic" rules like 30, 110 are not affected by noise, except for rule 22 which shows a slight increase of entropy. This marks the difference with respect to other stability indicators like the Lyapunov exponent, which is positive pr chaotic rules. 

In many case there is an entropy reduction by noise. This is particularly evident for rule 232. In this case, most of attractors are composed by configuration made by clusters of zeros and ones of at least width 2. 
These configurations are stable with respect to single perturbations inside a cluster but are ``connected’’ by perturbations at the boundary of a cluster. Due to the vanishing noise, these boundaries perform a self-annihilated random walk, so that the effect of the noise is to drive the system to the stable configurations $@0$ and $@1$. 

This is also the case of Rule 13 and 77, for which the asymptotic configuration is an alternation of zeros and ones, with occasionally clusters of double zeros which however can move and self-annihilate in the presence of noise. The same happens for instance to rule 58 for which a stable (translating) configuration is a repetition of $\{1,1,0\}$, and the noise serves to remove the $\{1,0,1,0\}$ defects.

For a few rules, entropy increases by noise, but not for the trivial effect of introducing disturbances, since the measure is always performed on the distribution of attractors. Let us take as an example rule 1. 

In this rule, the local pattern $\{0,0,0\}$ gives 1, all other patterns give zero. All isolated zeros in the initial configuration are removed, while isolated 1 maintains (every other step). So, the effect of noise is that of creating cluster of isolated ones in a greater number with respect to the random initialization process, and entropy increases.  

In this cases, the effect of noise is that of connecting basins with few states (with low statistical weight starting from a random configuration) to configurations belonging to the largest basin and thus increasing the entropy. 

The simpler, fixed-point dynamics of the Hopfield model, being more controllable, make it easier to understand entropy changes, especially in contrast to cellular automata, whose dynamics, with cycles of varying lengths, if often less clear.

Results for the Hopfield model are presented in Fig.~\ref{fig:Hopfield} for $L=17$ varying the number of attractors $K$. Specifically, for each $K$, we report the average values of $S$ and $S^*$ over 1000 repetitions. On average, $S$ is greater than $S^*$, though in a few cases, entropy increases when noise is introduced.

To better understand the nature of these two different behaviors, in the following we discuss two examples of minimal Hopfield network ($L=4$) where, respectively, a decrease and an increase of entropy is observed. The phase spaces are reported in Fig.~\ref{fig:HopfieldNoise}. 

In all networks, the same number of patterns ($K=6$) was stored according to equation \ref{eq:w}. 
In principle, there should be 12 attractors, corresponding to the stored patterns and their inverse, but only occasionally this happens (Fig.~\ref{fig:HopfieldIdeal}). Normally, there are fewer attractors, not corresponding to the stored patterns, as reported in Fig.~\ref{fig:Hopfield}.

Each attractor has a basin of transient states that converge to it (blue nodes). Since the dynamics of the model is "contracting", the entropy reduces from that of the initial distribution in which all system states are equiprobable (the dynamics is governed by Eq.~\eqref{eq:sigma} and represented by black arrows in the figure).

The effect of a vanishing noise is that of enabling transitions between attractors. Red arrows in Fig.~\ref{fig:HopfieldIdeal} and Fig.~\ref{fig:HopfieldNoise}  mark the elements of the transition matrix $W$. The noise redistribute the probability among the attractors and potentially change the entropy. In general, entropy decreases when certain attractors can only be exited due to noise, making them less populated, Fig.~\ref{fig:HopfieldNoise}(c)).

Conversely, when noise facilitates transitions between all attractors, entropy increases, as the system can explore a larger portion of state space, distributing probability more evenly among attractors, Fig.~\ref{fig:HopfieldNoise}(d)).

\section{Conclusions}

We investigated the effects of noise in the Hopfield model and elementary cellular automata, treating them as discrete dynamical systems. For both systems, we mapped the complete phase space, including accessible states, attractors, and their basins of attraction, for small lattice sizes (up to 
$L=17$). Starting from a maximal entropy distribution, where all configurations are equally probable, we demonstrated how the deterministic dynamics change this distribution, leading to entropy reduction.

Next, we examined how a vanishing noise alters the phase-space landscape by connecting attractors and modifying the asymptotic probability distribution over configurations. While in some cases the noise has little or none effect on the steady-state distribution, in most cases, it reduces entropy by funneling unstable attractors into the basins of more stable ones.

In rare cases, the opposite effect occurs, increasing the entropy. This phenomenon is linked to the instability of the largest attractors, which, under the effect of noise, become less populated in favor of smaller ones.

Exploring the relationship between attractor stability and noise-induced transitions is crucial, particularly in biological systems and neural networks, where noise plays a key role in learning, memory retrieval, and dynamical stability. In models such as Hopfield networks, noise helps prevent the system from being trapped in spurious states, thus improving retrieval performance. Future work will focus on investigating the connection between stability (Lyapunov exponents) and attractors, as well as the application of advanced statistical techniques to effectively handle larger lattices.


\section*{Declarations}

Some journals require declarations to be submitted in a standardised format. Please check the Instructions for Authors of the journal to which you are submitting to see if you need to complete this section. If yes, your manuscript must contain the following sections under the heading `Declarations':

\section*{Funding} This publication was produced with the co-funding of European Union - Next Generation EU, in the context of The National Recovery and Resilience Plan, Investment 1.5 Ecosystems of Innovation, Project Tuscany Health Ecosystem (THE), CUP: B83C22003920001.
\section{Conflict of interest/Competing interests} Not Applicable
\section{Ethics approval and consent to participate} Not Applicable
\section{Consent for publication} Not Applicable
\section{Data availability} Not Applicable
\section{Materials availability} Not Applicable
\section{Code availability} The simulation code is available upon request from authors. 
\section{Author contribution} All authors contributed equally. 

\bibliography{ECA}
\end{document}